\documentclass[onecolumn]{aastex63}
\usepackage[figuresright]{rotating}
\usepackage{graphicx}
\DeclareGraphicsExtensions{.pdf,.jpg,.png}
\usepackage{float}

\usepackage{threeparttable}

\begin{document}


\title{Resolving the Nuclear Radio Emission from M32 with Very Large Array}
\email{sijiapeng@smail.nju.edu.cn; lizy@nju.edu.cn}



\author{Sijia Peng}
\affiliation{School of Astronomy and Space Science, Nanjing University, Nanjing 210023, China}
\affiliation{Key Laboratory of Modern Astronomy and Astrophysics, Nanjing University, Nanjing 210023, China}

\author{Zhiyuan Li}
\affiliation{School of Astronomy and Space Science, Nanjing University, Nanjing 210023, China}
\affiliation{Key Laboratory of Modern Astronomy and Astrophysics, Nanjing University, Nanjing 210023, China}

\author{Lor\'{a}nt O. Sjouwerman}
\affiliation{National Radio Astronomy Observatory, Socorro, NM 87801, USA}

\author{Yang Yang}
\affiliation{Shanghai Astronomical Observatory, Chinese Academy of Sciences, Shanghai 200030, China}
\affiliation{Key Laboratory of Radio Astronomy, Chinese Academy of Sciences, China}

\author{Fuguo Xie}
\affiliation{Shanghai Astronomical Observatory, Chinese Academy of Sciences, Shanghai 200030, China}
\affiliation{Key Laboratory for Research in Galaxies and Cosmology, Chinese Academy of Sciences, China}

\author{Feng Yuan}
\affiliation{Shanghai Astronomical Observatory, Chinese Academy of Sciences, Shanghai 200030, China}
\affiliation{Key Laboratory for Research in Galaxies and Cosmology, Chinese Academy of Sciences, China}

\begin{abstract}
The Local Group dwarf elliptical galaxy M32 hosts one of the nearest and most under-luminous super-massive black holes (SMBHs) ever known, offering a rare opportunity to study the physics of accreting SMBHs at the most quiescent state. 
Recent Very Large Array (VLA) observations have detected a radio source at the nucleus of M32, which is suggested to be the radio counterpart of the SMBH. 
To further investigate the radio properties of this nuclear source, we have conducted follow-up, high-resolution VLA observations in four epochs between 2015--2017, each with dual frequencies. 
At 6 GHz, the nuclear source is resolved under an angular resolution of $\sim$0\farcs4, exhibiting a coreless, slightly lopsided morphology with a detectable extent of $\sim$$2\farcs5$ ($\sim$10 parsec). 
No significant variability can be found among the four epochs. 
At 15 GHz, no significant emission can be detected within the same region, pointing to a steep intrinsic radio spectrum (with a 3\,$\sigma$ upper limit of -1.46 for the spectral index).
We discuss possible scenarios for the nature of this nuclear source and conclude that a stellar origin, in particular planetary nebulae, X-ray binaries, supernova remnants or diffuse ionized gas powered by massive stars, can be ruled out. 
Instead, the observed radio properties can be explained by synchrotron radiation from a hypothetical wind driven by the weakly accreting SMBH. 

\end{abstract}




\section{Introduction} 

Multi-wavelength, wide-field surveys in the past decades have revealed that most super-massive black holes (SMBHs) at low redshift are radiatively quiescent \citep{2008ARA&A..46..475H}, manifesting themselves as low-luminosity active galactic nuclei (LLAGNs).
As accretion rates drop from close to the Eddington limit to far below it, the radiative energy output (i.e., {\it feedback}) of SMBHs is thought to become less important compared to mechanical energy output \citep{2012ARA&A..50..455F, 2013ARA&A..51..511K}, which is mediated by outflows (in the form of jets and/or winds) powered by an advection-dominated accretion flow (ADAF; \citealp{2014ARA&A..52..529Y}).
Understanding the detailed physical working of LLAGNs, in particular their feedback modes, is a vital part of our understanding of the interaction and co-evolution between SMBHs and their host galaxies over cosmic time.

Finding direct evidence for LLAGN feedback, however,  proves to be a challenging task.  
Currently the most compelling evidence connecting large-scale mechanical power with a LLAGN comes from the observation of radio bubbles (often spatially coincident with X-ray cavities) inflated by relativistic jets, which are typically found in massive elliptical galaxies, galaxy groups and galaxy clusters \citep{2012NJPh...14e5023M}.
Whether this feedback mode is prevalent in lower-mass systems is difficult to assert, since the mechanical power from the SMBH is not easily dissipated in low density environments.  
A particularly acute case is found with our own Milky Way galaxy, which harbors the nearest SMBH, commonly known as Sgr A* \citep{2001ARA&A..39..309M}.
On the one hand, Sgr A* has been the prime testbed for the theories of SMBH accretion and occupied a prominent position in the development of the ADAF-jet paradigm \citep{1998ApJ...492..554N, 2000A&A...362..113F,2002A&A...383..854Y}.
On the other hand, despite its virtue of proximity, Sgr~A* is persistently seen as a compact radio source down to the vicinity of its event horizon \citep{2018ApJ...859...60L}, disfavoring the presence of an elongated jet.
Observational searches for a putative jet from Sgr A* on parsec- to kiloparsec-scales also remain inconclusive (\citealp{2013ApJ...779..154L, 2019ApJ...875...44Z}, and discussions therein).
Clearly, our physical understanding of LLAGNs, in particular Sgr A*, will not be complete without a census of SMBHs at the most quiescent state.

As an intriguing coincidence, lying at the low-luminosity end of all known SMBHs are the three nearest ones, respectively hosted by the three Local Group galaxies: Milky Way, M31 and M32.
As the nearest example of a compact elliptical galaxy, M32 harbors a SMBH with a mass of $\sim$$\rm 2.5\times 10^{6}~M_{\odot}$ estimated from stellar dynamical measurements \citep{2010MNRAS.401.1770V, 2002MNRAS.335..517V}.
An X-ray source was found to be coincident with the nucleus of M32 based on {\it Chandra} observations \citep{2003ApJ...589..783H}. 
While this source exhibits a power-law X-ray spectrum (with a photon-index of $\sim$2.4) characteristic of LLAGNs, its 2--10 keV intrinsic luminosity is only $\sim$$10^{36}{\rm~erg~s^{-1}}$, or $\sim$$10^{-8.5}$ of the Eddington luminosity of the putative SMBH \citep[hereafter Y15]{2015ApJ...807L..19Y}.
Using NSF's Karl G. Jansky Very Large Array (VLA) observations at a central frequency of 6.6 GHz with a bandwidth of 2 GHz, Y15 detected a weak radio source (with a luminosity of ${\nu}L_{\nu}$[6.6 GHz] $\sim 2\times10^{32}{\rm~erg~s^{-1}}$) coincident with the optical nucleus of M32, which they argued is most likely the radio counterpart of the SMBH.
Together with Sgr A* and M31* \citep{1992ApJ...390L...9C, 2017ApJ...845..140Y}, the LLAGN in M32 provides a rare but promising opportunity to study feedback from the most under-luminous SMBHs on parsec-scales. 

In this work, we present follow-up high-resolution, dual-band VLA observations to further explore the nuclear radio source in M32.
Section 2 describes our new VLA observations and data reduction procedures. Section 3 presents the main observational results, in particular morphological and spectral properties of the nuclear source. Section 4 is devoted to understanding the nature of this source.
To be consistent with previous work, we adopt a distance of 780 kpc for M32 ($1\arcsec$ corresponds to 3.8 pc; \citealp{2008ChJAA...8..369Y}). Quoted errors are at 1\,$\sigma$ confidence level, unless otherwise stated.

\section{Observations and Data Preparation} {\label{sec:data}

Our VLA A-configuration observations of M32 were initially scheduled for the 2015A semester (Project ID: 15A-049, PI: L. Sjouwerman). However, due to a failure in the antenna control system, only one of the three requested observing epochs was successfully conducted, which resulted in two 3-hr integrations on 2015-7-20, one in C-band and the other in Ku-band. 
Subsequently, on 2016-12-13, 2016-12-14 and 2017-1-22, we conducted three additional epochs of dual-band observations through a new program (Project ID: 16B-073) with the same setting as in 15A-049. In total, the on-source integration time is about 6.5 (7.6) hours in C-band (Ku-band). In all four epochs, C-band (Ku-band) observation was centered at 6.0 (15.0) GHz, with a bandwidth of 4 (6) GHz as captured with the 3-bit samplers. Using a visibility integration time of 2 seconds, the available bandwidth was divided in 1 MHz dual polarization channels to aid the selective removal of RFI.
Phase-referenced mode was adopted, with J0038+4137 serving as phase calibrator 
every 10 minutes for C-band and every 8 minutes for Ku-band.
3C48\footnote{3C48 underwent a major event since 2016, however, it did not significantly affect our data analysis and result (private communication with NRAO helpdesk). See also https://science.nrao.edu/facilities/vla/docs/manuals/obsguide/modes/pol} was used as flux and bandpass calibrator with a $\rm 3\%$ systematic uncertainty \citep{2013ApJS..204...19P}.  
In total, we obtained 8.5 hours on-source integration for C-band and 10 hours for Ku-band.
A log of the VLA observations are given in Table \ref{tab:infor}.

The data were calibrated using the Common Astronomy Software Applications\footnote{https://casa.nrao.edu} (CASA, version:4.6.0-REL) in standard procedures\footnote{https://casaguides.nrao.edu/index.php/VLA\_Continuum\_Tutorial\_3C391-CASA4.6}. We manually flagged bad data due to radio frequency interference and other instrumental effects. Antenna position corrections, flux density scaling, delay, bandpass, gain and phase calibrations were applied. 
We have examined potential flux variability in the sources of interest, in particular the nuclear source, but found that the flux densities measured in individual epochs, albeit of substantial uncertainties, were consistent with each other. 
Subsequently, using task CONCAT, we combined the target visibilities of all four epochs to yield a concatenated data. 
The final image was produced for each band by applying task CLEAN on the concatenated data under natural weighting, without self-calibration because of the weak signal.
Below we shall focus on this concatenated image. 
The Ku-band image has a synthesized beam size of $\rm 0\farcs15\times0\farcs14$ and a rms noise of  $\rm 1.3\  \mu Jy\ beam^{-1}$,
while the C-band image has a synthesized beam size of $\rm 0\farcs40\times0\farcs37$ and a rms noise of $\rm 1.0\  \mu Jy\ beam^{-1}$. 
For comparison, the VLA B-configuration C-band image presented in Y15 achieved a resolution of $\rm 1\farcs14\times1\farcs09$ and a rms noise of  $\rm 1.2\  \mu Jy\ beam^{-1}$. 

Following Y15, we registered the astrometry between our VLA images with an optical image of M32, to accurately determine the putative position of the SMBH on the radio images. To do so, we relied on four planetary nebulae (PNe) that are commonly detected in the VLA images and an archival {\it Hubble\ Space\ Telescope} WFC/F502N image (HST Program ID: 11714; Figure \ref{fig:PN}). The latter image, with a resolution of $\rm 0\farcs04$, is sensitive to the $\left[\rm O\  \uppercase\expandafter{\romannumeral3}\right]$ forbidden line of PNe. 
The basic information of the four PNe, designated R2--R5, are listed in Table \ref{tab:PN}. 
We note that R5 was not detected in the B-configuration image of Y15, but has been identified as a PN by optical spectroscopic observations \citep{1989ApJ...339...53C, 2011MNRAS.415.2832S}. 
We registered the radio and optical images by matching the PN centroids, taking into account the positional uncertainties in the respective images. 
The optical nucleus of M32 (i.e., the intensity peak) is then found at $[\rm R.A. , Dec$] = $[00^{h}42^{m}41\fs836, +40\arcdeg51\arcmin54\farcs99]$ in the radio image, with a 95\% error of $\rm 0\farcs15$ on R.A. and $\rm 0\farcs14$ on Dec., shown as the magenta cross in Figure \ref{fig:M32}. 
Here the quoted error takes in account both the radio and optical positional uncertainties.

\section{Results} 
The innermost $6''$$\times$$6''$ region of M32 at 6 GHz is shown in Figure \ref{fig:M32}a.
The black contours are at levels of $\rm (-3, 3)\times RMS$ corresponding to $\rm (-3.0,3.0)\  \mu Jy\ beam^{-1}$ and the negative components are displayed in dashed lines. 
At 6 GHz, significant emission is found around the optical nucleus and exhibits an extended morphology, i.e., significantly broader than a point source, which was hinted in the B-configuration image of Y15 (reproduced as cyan contours) and is now confirmed under higher resolution.
We emphasize that the extended structure is not an artifact due to insufficient calibration, since the point-like PNe argue against such a case.
Meanwhile, there are only a few negative 3\,$\sigma$ components, showing that the image was not overly cleaned. 

Somewhat surprisingly, there is no sign of enhanced emission at the putative position of the SMBH (marked by a `+' sign in Figure \ref{fig:M32}, as determined in Section \ref{sec:data}), as would be expected from an unresolved radio core, although a 3\,$\sigma$ peak is marginally coincident with that position. A few other $\gtrsim$3\,$\sigma$ peaks are present to the southeast of the nucleus, reaching a projected distance of $\sim$$1\farcs3$ ($\sim$5 pc), whereas emission from the western side is slightly weaker and less extended. 
Summing up the $\rm 3\,\sigma$ peaks we find a total flux density of $\sim$$\rm 17~{\mu}Jy$.
The 6 GHz integrated flux density within a $\rm 1\farcs3$-radius circle centering at the optical nucleus is measured to be $\rm 25.1\pm5.8\ \mu Jy$, which is consistent with the value of $\rm \sim23\ \mu Jy$ measured from the same region in the B-configuration image (at a slightly different central frequency of 6.6 GHz). 
The flux densities of the nuclear source measured in the concatenated image as well as in the individual epochs are listed in Table~\ref{tab:infor}.
We note that Y15 reported a nearly twice higher integrated intensity based on an elliptical Gaussian fit to a larger aperture. 
This suggests that the extended nuclear emission is partially resolved out under A-configuration. 
Alternatively, this might be due to intrinsic flux variation of the nuclear source on a timescale of 1.5 years. 
Significant flux variation in the nuclear X-ray emission was found over timescales of a few days to $\sim$11 years (Y15).
However, the fact that the radio source has an extent of a few parsecs argues against a substantial variability.

To enhance the visualization of the extended structure, we re-image the C-band concentrated data to obtain a tapered image, setting CLEAN parameters $\rm OUTERTAPER = 150\ k\lambda$ and $\rm RESTOREBEAM=1\farcs09$, which enforces tapering on outer baselines in the {\it uv-plane} and producing Gaussian restoring beam. This results in a tapered image with a resolution nearly identical to the B-configuration image. The corresponding intensity contours at levels of (-3, 3, 3.8)$\times$ RMS (= $\rm 1.4\  \mu Jy\ beam^{-1}$) are shown in Figure \ref{fig:M32}a (red curves).
As expected, the tapered image appears smoother than the untapered one, but otherwise preserves the overall morphology of the nuclear emission. 

At 15 GHz (Ku-band), there is no significant nuclear emission in the concentrated image nor in the image of any single epoch. Therefore, we can only obtain a $\rm 3\,\sigma$ upper limit of $\rm 6.6\  \mu Jy$, which is derived from a tapered Ku-band image at a resolution equal to that of the C-band image.
Defining a spectral index $\alpha$ as $S_{\nu} \propto \nu^{\alpha}$ and adopting a nominal 6 GHz integrated flux density of $\rm 25.1\ \mu Jy$, we find the corresponding 3\,$\sigma$ upper limit $\alpha < -1.46$ for the nuclear emission in M32. 
We note that this is a conservative estimate; adopting the integrated intensity of $\sim$47 $\mu$Jy (at 6.6 GHz) measured from the B-configuration image (Y15) would result in an even more negative value of $\alpha$. 
The total luminosity over the range of 1--20 GHz is $\lesssim 10^{33}{\rm~erg~s^{-1}}$. 
Figure~\ref{fig:sed} displays the observed radio flux densities at 6.0 GHz and 6.6 GHz, as well as $\rm 3\,\sigma$ upper limits a 8.4 GHz (from \citealp{2003ApJ...589..783H}) and 15 GHz. 
It is evident that the extended nuclear source exhibits a rather steep spectrum over 6--15 GHz. 
For comparison, the observed 6 and 15 GHz flux densities of the four PNe (Table~\ref{tab:PN}) are also shown as open circles in Figure~\ref{fig:sed}. Except for R2, these PNe show a flat spectrum, presumably due to free-free emission that is characteristic of radio PNe.  



\section{Discussion and Conclusion}

The VLA A-configuration observations presented above provide new information about the radio emission from the nucleus of M32, where one of the nearest and most under-luminous SMBHs is located.  
These high resolution observations resolve the nuclear emission at 6 GHz, which has a detectable size of $\sim$10 parsecs.
Interestingly, there is no definite sign of an intensity peak at the putative position of the SMBH, as would be expected from a self-absorbed core often found in LLAGNs, including in particular the well-known case of Sgr A*. 
There is evidence that the nuclear source (hereafter referred to as R1 following Y15, for clarity) is lopsided, with the southeastern side slightly more extended than the northwestern side, but the overall morphology does not resemble a classical collimated jet-like structure (Figure~\ref{fig:M32}a).   
Despite the unprecedented sensitivity afforded by our observations, no significant emission is detected at 15 GHz, revealing that R1 has a steep spectrum over the 6--15 GHz range (spectral index $\alpha < -1.46$). 
These observed properties place tight constraints on the origin of the radio emission, which in principle can be attributed to i) discrete stellar objects, (ii) diffuse circumnuclear medium powered by stellar activity, or iii) an outflow driven by the weakly accreting SMBH. We shall consider more specific possibilities below.

The stellar content of M32 consists of two old populations: $\rm \sim40\% \pm17\%$ of the stellar mass in a 2-5 Gyr-old, metal-rich population and the remaining $\rm \sim55\% \pm 21\%$ in stars older than 5 Gyr, with slightly sub-solar metallicities \citep{2012ApJ...745...97M}. 
The observed radio flux may be contributed, partially or predominantly, by stellar objects. We have detected in our radio images four PNe, indicating that PNe are among the brightest stellar sources at these frequencies. If the observed radio emission from within the central arcsecond, in particular the 3\,$\sigma$ intensity peaks, are due to unidentified PNe, their optical counterparts should be readily detected through the [O III] line. This however is inconsistent with the non-detection by the SAURON integral-field spectroscopic imaging \citep{2011MNRAS.415.2832S}. 

X-ray binaries (XRBs) involving an accreting neutron star or stellar-mass black hole, which are certainly present in proportion to the old stellar populations of M32 \citep{2007A&A...473..783R}, may also account for the observed radio flux as well as the persistent nuclear X-ray source (designated X1; \citealp{2003ApJ...589..783H}) as seen by {\it Chandra} and {\it XMM-Newton} (Y15).  
However, a single XRB is unlikely to be detected by our VLA images. 
To see this, we shall recall that X1 has a mean 2--10 keV luminosity of $\sim 10^{36}{\rm~erg~s^{-1}}$ (Y15), which implies that the putative XRB is accreting at the low state, whose X-ray and radio emission would follow the empirical fundamental plane relation of black hole activity (hereafter the FP; \citealp{2003MNRAS.345.1057M, 2004A&A...414..895F}).
In this case, the expected 6 GHz luminosity of X1 is $\lesssim 10^{30}{\rm~erg~s^{-1}}$ (or a flux density $\lesssim 0.2{\rm~{\mu}Jy}$), which falls far below the sensitivity of our C-band image\footnote{While this argument is based on black hole binaries, it also applies for neutron star binaries, which have on average ten times lower radio luminosity than black hole binaries at the same X-ray luminosity \citep{2006MNRAS.366...79M}}. 
Indeed, even the bright X-ray source, X3 \citep{2003ApJ...589..783H}, located at an off-nucleus distance of $\sim$$9''$ with a $\sim$80 times higher X-ray flux than X1, is undetected in our deep C-band images. 
We note that a transient radio source, similar to those found in the Galactic center and likely tracing outburst from a black hole binary (e.g., \citealp{1992Sci...255.1538Z}), was considered by Y15 a possible explanation for R1. 
Our new observations now clearly show that R1 is a persistent source, thus the possibility of a transient nuclear source can be safely ruled out.

R1 with its 10-parsec extent and the apparent absence of a dominant core, may be interpreted as an evolved supernova remnant (SNR), although in such a case X1 will be irrelevant, because the latter exhibits significant variability over a decade (Y15).   
The Galactic center SNR Sgr A East is the best-known example of a SNR found in the close vicinity of a SMBH \citep{1996A&ARv...7..289M}.
When put at the distance of M32, Sgr A East can mimic the observed size and morphology of R1.
For its physical size, the radio luminosity of R1 is also marginally compatible with the luminosity-diameter relation of shell-like SNRs  \citep{1998ApJ...504..761C}. 
However, at present the nuclear environment of M32 strongly disfavors the birth of supernovae (SNe).
First, there is no evidence for recent or on-going star formation across the entire galaxy of M32 \citep{2012ApJ...745...97M}, and massive stars are virtually absent in the nucleus \citep{2000ApJ...532..308B}.
This precludes the occurrence of core-collapse SNe in the past few million years at least.
While the old stellar populations can give rise to Type Ia SNe, the probability of having one SN\,Ia within the central 5 parsec of M32 in the past $10^3$ yr is only about 0.1\%. 
To obtain this estimate, we have assumed an empirical SN\,Ia birth rate of $0.044$ per century per $10^{10}{\rm~M_\odot}$ for elliptical/lenticular galaxies \citep{2005A&A...433..807M} and an underlying stellar mass of $2\times10^7{\rm~M_\odot}$ in the M32 nucleus \citep {1998AJ....116.2263L}. 
While the exceedingly high stellar density of the nucleus may enhance dynamical formation of close binaries and hence the birth rate of SNe Ia (e.g., \citealp{2019ApJ...878...58S}), it is unlikely that this mechanism can lead to more than a ten-fold increase, so as to be compatible with surveys of SNe Ia in the local universe. 
In addition, the spectral index of R1 is significantly steeper than that of the known radio SNRs in the Galaxy and M33, which have a typical value of $\rm -0.5$ \citep{2014BASI...42...47G, 2019ApJS..241...37W}.
Therefore, we conclude that R1 is highly unlikely to be the radio counterpart of a SNR. 

On the basis of the absence of nuclear star formation, Y15 argued that the observed radio flux cannot be dominated by free-free emission from diffuse ionized gas, as in the case of Sgr A West, an ionized gas structure within the inner few parsecs of the Galactic center, which is thought to be powered by massive stars \citep{1996A&ARv...7..289M}. 
The observed steep spectrum over 6--15 GHz reinforces this assessment. In particular, the 3\,$\sigma$ upper limit at 15 GHz constrains the contribution from free-free emission to less than $\sim$30\% at 6 GHz, assuming a canonical electron temperature of $10^4$ K \citep{2011ApJ...727...35D}. 
Finally, adopting an average 6 GHz flux density of $\rm 4\times10^{7}~Jy$ from the active Sun \citep{2005psci.book.....A} and an enclosed stellar mass of $2\times10^7{\rm~M_\odot}$ \citep{1998AJ....116.2263L}, the collective 6 GHz flux density of the old stars in the M32 nucleus can be estimated to be merely $\rm 0.03\ \mu Jy$, far less than the observed value. 

The above estimates strongly disfavor a stellar origin for R1.  Thus it is more likely that R1 is associated with the central SMBH. 
In this case, the anticipated radio feature would be a collimated jet or a jet plus lobe structure. 
Recent high-resolution radio interferometric observations provide growing evidence for the existence of parsec-scale jets from LLAGNs \citep{2014ApJ...787...62M}.
The steep spectrum of R1 is also consistent with jet synchrotron. 
However, a collimated morphology is not seen in R1.  Moreover, the absence of a dominant core is rarely seen in LLAGNs (a possible exception might have been found in NGC\,404; \citealp{2014ApJ...791....2P}), and in fact, is in marked contrast with the cases of Sgr A* and M31* \citep{2017ApJ...845..140Y}, the only other two SMBHs known to have extremely low Eddington ratios. We note that in Sgr A* the compact core emission below 40 GHz is produced by non-thermal electrons in the inner region of the ADAF \citep{2003ApJ...598..301Y}, which exhibits a spectra index of $0.28 \pm 0.03$ over 1--40 GHz \citep{2015ApJ...802...69B}, much flatter than that of M32 R1. 
The non-detection of a radio core in M32 may imply that the non-thermal electrons are rapidly suppressed as the system brightens (i.e., the SMBH of M32 has $\gtrsim10$ times higher Eddington ratio than Sgr A*), due to the efficient synchrotron boiler effect (e.g., \citealp{2009MNRAS.392..570M}).

On the basis of its unusual morphology and steep spectrum, we suggest that R1 traces the synchrotron radiation from a wind launched from the accretion flow around the SMBH. 
Known as the ADAF wind, such an energetic outflow is a generic prediction by theories and numerical simulations of hot accretion flows in LLAGNs \citep{1999MNRAS.303L...1B,2012ApJ...761..129Y, 2012ApJ...761..130Y, 2015ApJ...804..101Y}. 
Originating from the corona region of the hot accretion flow, the ADAF wind would have a wide opening angle as compared to the highly collimated jet, thus more naturally matching the observed radio morphology.  
Relativistic electrons responsible for the synchrotron may be produced in the corona of ADAF, where frequent magnetic reconnection is expected to occur \citep{2009MNRAS.395.2183Y}. 
Alternatively, relativistic electrons may be produced via the outward propagating wind shocking the circumnuclear medium over a region of 10 parsec, corresponding to $\sim$$4\times10^{7}R_s$ (here $R_s \approx 2.4\times10^{-7}\rm~pc$ is the Schwarzschild radius of the putative SMBH). 
To our knowledge, there has been no dedicated theoretical effort to predict the observational signatures of an ADAF wind on relevant physical scales. 
Here we shall provide a simple estimate of the energetics related to the radio emission. 
First we notice that the radiative efficiency of an ADAF ($\epsilon \equiv L_{\rm bol}/\dot{M}_{\rm net}c^2$) at highly sub-Eddington rates scales roughly with the net mass accretion rate (i.e., the mass inflow rate into the event horizon) as $\epsilon \propto (\dot{M}_{\rm net}/\dot{M}_{\rm Edd})^{0.7} $ \citep{2012MNRAS.427.1580X}, where $L_{\rm bol}$ is the bolometric luminosity of the SMBH and $\dot{M}_{\rm Edd}c^2 \equiv 1.3\times10^{39}(M_{\rm BH}/M_\odot){\rm~erg~s^{-1}}$ is the Eddington accretion rate. 
Assuming that the 2--10 keV luminosity of X1 ($10^{36}{\rm~erg~s^{-1}}$) is 10\% of the bolometric luminosity, and adopting a canonical value of $\delta = 0.5$, the fractional viscous heating of electrons in the ADAF\footnote{See \citet{2012MNRAS.427.1580X} for discussions about this parameter; adopting a smaller value of $\delta$ would result in a higher $\dot{M}_{\rm net}$.}, we may infer $\dot{M}_{\rm net}c^2 \approx 8.0\times10^{-7}\dot{M}_{\rm Edd}c^2 \approx 2.6\times10^{39}{\rm~erg~s^{-1}}$. 
Numerical simulations show that the ADAF wind carries a kinetic energy of $10^{-3}\dot{M}_{\rm net}c^2$ \citep{2015ApJ...804..101Y}, which is then $\sim2.6\times10^{36}{\rm~erg~s^{-1}}$
according to the above estimate.
This value is sufficient to account for the observed synchrotron radiation, given that typically 1\% of the kinetic energy goes to shock-accelerating relativistic electrons \citep{2015MNRAS.447.3612N}. 
The ADAF wind may also have produced soft X-ray emission upon shock-heating the circumnuclear medium. 
We note that the observed 0.5--2 keV luminosity from X1 ($\lesssim10^{36}{\rm~erg~s^{-1}}$; Y15) is compatible with the estimated wind kinetic energy. 

Given the above considerations, we suggest that the radio emission from the nucleus of M32 is best explained by a hypothetical wind symbiotic to the highly sub-Eddington accretion onto the SMBH. 
A more definitive view of the interaction between this wind and the circumnuclear medium, both in M32 and in other LLAGNs, invites future high-sensitivity observations afforded by the Next Generation Very Large Array and the Square Kilometre Array.

\acknowledgments
The National Radio Astronomy Observatory is a facility of the National Science Foundation operated under cooperative agreement by Associated Universities, Inc. 
We thank Zhiyu Zhang for helpful discussions.
S.P and Z.L. acknowledge support by the National Key Research and Development Program of China (2017YFA0402703) and National Natural Science Foundation of China (grants 11873028, 11473010). 
FGX is supported in part by the National Key Research and Development Program of China (Grant No. 2016YFA0400804), the Natural Science Foundation of China (grant 11873074),
and the Youth Innovation Promotion Association of CAS (id 2016243).
FY is supported in part by the National Key Research and Development Program of China (Grant No. 2016YFA0400704) and the Natural Science Foundation of China (grant 11633006).

\clearpage

\begin{deluxetable*}{ccccccc}
\tablecaption{Log of VLA observations\label{tab:infor}}
\tablecolumns{7}
\tablenum{1}
\tablewidth{0pt}
\tablehead{
\colhead{Date} & \colhead{Frequency} & \colhead{Beam size} &\colhead{RMS} &\multicolumn{2}{c}{Centroid position} & \colhead{R1 flux density} \\
\colhead{(yy-mm-dd)} & \colhead{(GHz)} & \colhead{$(\rm '',\ '',\ ^{o})$} &\colhead{$(\rm \mu Jy~beam^{-1})$} & \colhead{R.A.} & \colhead{Dec.} & \colhead{($\rm \mu Jy$)}  
}
\colnumbers
\startdata
2015-7-20 & 6.0 & $ 0.45, 0.38, -61.4$ & 2.2 & 00h42m41.836s & +40d51m54.99s & $16.3\pm7.9$ \\
 & 15.0 & $0.15, 0.14, -57.1$ & 2.6 & - & - &- \\
2016-12-13 & 6.0 & $ 0.39, 0.38, -76.4$ & 1.8 & 00h42m41.836s & +40d51m54.99s & $27.7\pm11.7$ \\ 
 & 15.0 & $0.15, 0.13, -0.4$ & 2.1 & - & - &- \\
2016-12-14 & 6.0 & $ 0.39, 0.36, -89.0$ & 1.8 & 00h42m41.836s & +40d51m54.99s & $18.5\pm8.3$ \\
 & 15.0 & $ 0.16, 0.14, 42.5$ & 2.9 & - & - &- \\
2017-1-22 & 6.0 & $ 0.42, 0.38, -76.1$ & 2.1 & 00h42m41.836s & +40d51m54.99s & $32.7\pm10.2$ \\
 & 15.0 & $ 0.17, 0.14, 27.1$ & 2.6 & - & - &- \\
\hline
Concatenated & 6.0 & $ 0.41, 0.37, -75.3$ & 1.0 & 00h42m41.836s & +40d51m54.99s & $25.1\pm5.8$ \\
& 15.0 & $0.15, 0.14, 21.9$ & 1.3 &  - &- & $\leq 6.6$ \\
\enddata
\tablecomments{(1) Observation date. (2) Central frequency. (3) Synthesized beam FWHM of major and minor axes, and position angle of the major axis. (4) Image rms level. (5)-(6) Coordinates of the source centroid, determined according to the optical nucleus and same for all images. (7) The flux density of R1, measured from IMFIT.}
\end{deluxetable*}

\begin{deluxetable*}{cccccccccc}
\tablecaption{Log of PNe \label{tab:PN}}
\tablecolumns{6}
\tablenum{2}
\tablewidth{0pt}
\tablehead{
\colhead{Name} & \multicolumn{2}{c}{Centroid position } & \multicolumn{2}{c}{Offset} & \colhead{$\rm \Delta p$} & \multicolumn{3}{c}{Flux density ($\rm \mu Jy$)}  & \colhead{Spectral index}  \\
\colhead{} & \colhead{R.A.} & \colhead{Dec.} &\colhead{R.A.} & \colhead{Dec.} & \colhead{} & \colhead{6 GHz} & \colhead{6.6 GHz} & \colhead{15 GHz} 
}
\colnumbers
\startdata
R2 & 00h42m41.889s & +40d51m58.63s & $\rm 0.05$ & $\rm 0.31$ & $\rm 0.05$ & $\rm 13.8 \pm2.9$  & $\rm 10.5\pm 0.7$  & $\rm 5.6\pm0.9$ & $\rm -0.98\pm0.29$\\
R3 &00h42m42.541s & +40d51m56.61s & $\rm 0.06$ & $\rm 0.19$ & $\rm 0.08$  & $\rm 3.8 \pm1.0$  & $\rm 8.1\pm 0.2$  & $\rm 5.4\pm0.3$  & $\rm 0.38\pm0.29$ \\
R4 & 00h42m42.663s & +40d51m56.62s & $\rm 0.18$& $\rm 0.28$ & $\rm 0.05$ &  $\rm 10.0 \pm1.6$  & $\rm 8.2\pm 0.2$  & $\rm 11.4\pm3.3$  & $\rm 0.14\pm0.36$ \\
R5 & 00h42m42.248s & +40d51m49.17s  & $\rm 0.05$ & $\rm 0.22$  & $\rm 0.09$ & $\rm 3.2\pm 0.6$  & $ - $  & $\rm 4.9\pm0.4$ & $\rm 0.47\pm0.22$ \\
\enddata

\tablecomments{(1) PN name. (2)-(3) Coordinates of PN centroid in C-band. (4)-(5)  Right ascension and Declination offset between optical and C-band images, in units of arcsecond. 
(6) Radio positional uncertainty, defined as  $\rm \Delta p= 60~mas \times (100~GHz/FREQ) \times (10~km/BSL)/SNR $, in units of arcsecond. Here, FREQ denotes the central frequency, BSL denotes the length of the longest baseline and SNR denotes the signal-to-noise ratio of the peak emission. (7)-(9) 6 GHz, 6.6 GHz and 15 GHz flux densities. The 6.6 GHz flux density is measured from the B-configuration image (Y15). All flux densities and errors are based on IMFIT and corrected for the primary beam. (10) Spectral index measured between 6 and 15 GHz.}
\end{deluxetable*}

\clearpage
\begin{figure}[h]
\centering
\includegraphics[width=1\textwidth]{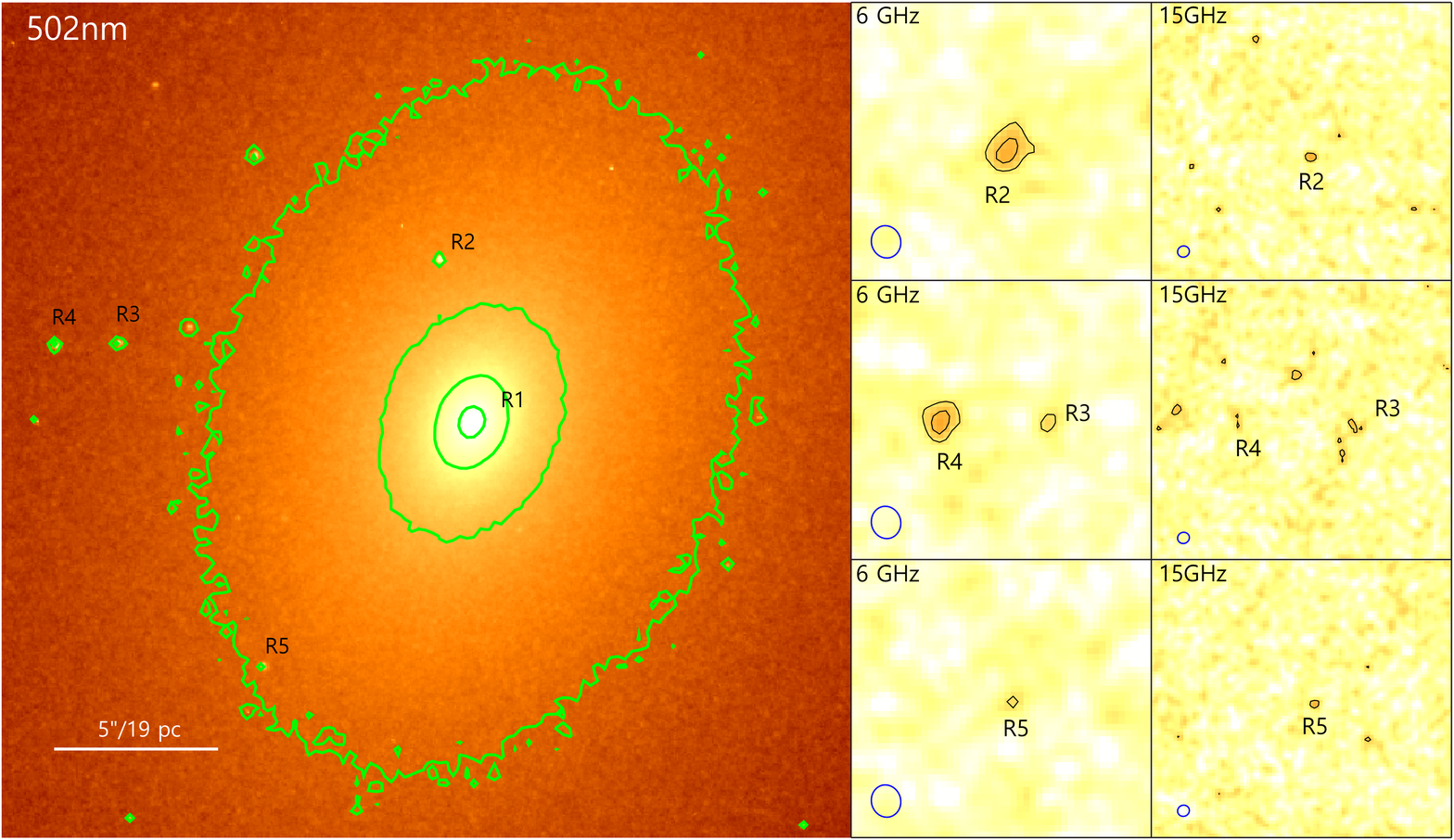}
\caption{\emph{Left}:  {\it HST}/WFC3/F502N image showing the bulk stellar content of M32 as well as PNe (denoted by R2, R3, R4, and R5) traced by their $\left[\rm O\uppercase\expandafter{\romannumeral3}\right]$ emission line. R1 marks the position of the nuclear radio source. Also plotted are selected intensity contours of the same image. \emph{Right}: A close-up of R2--R5 as seen in the VLA 6 and 15 GHz images. The black contours are at levels of $\rm (3, 5)\times RMS$ in the respective clean images. 
The synthesis beam is marked by a blue ellipse at the \textbf{corner} of each panel. \label{fig:PN}} 
\end{figure}

\begin{figure}
\centering
\includegraphics[width=1\textwidth]{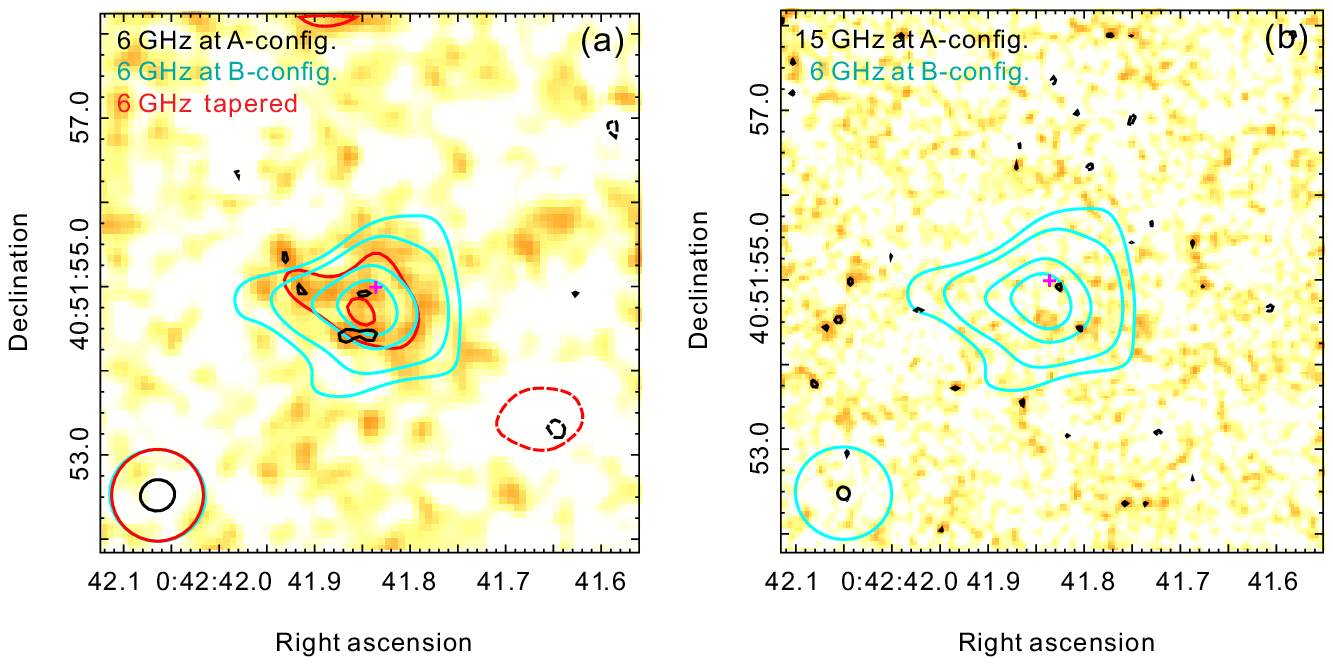}
\caption{
(a): VLA A-configuration 6 GHz image of the inner $6''\times6''$ region. The black contours are at levels of $\rm (-3, 3)\times RMS$ (=$\rm 1.0\  \mu Jy\ beam^{-1}$), with the negative components shown as dashed lines. The red contours are at levels of $\rm (-3, 3, 3.8)\times RMS$ (=$\rm 1.4\  \mu Jy\ beam^{-1}$) of the tapered 6 GHz image.  The cyan contours are at levels of $\rm (4, 5, 7, 8)\times RMS$  (=$\rm 1.2\  \mu Jy\ beam^{-1}$) in the VLA B-configuration 6.6 GHz image duplicated from Y15. (b): 15 GHz image of the same region as in (a). The black contours are at levels of $\rm (-3, 3)\times RMS$  (=$\rm 1.3\  \mu Jy\ beam^{-1}$) with negative components shown as dashed lines. 
The cyan contours are the same as in (a). 
In both panels, the synthesized beams of the respective images are indicated by the color-coded ellipses at the lower left corner, while the optical nucleus of M32 is marked by a magenta cross whose size represents the 95\% positional uncertainty. 
\label{fig:M32} }
\end{figure}

\begin{figure}[h]
\centering
\includegraphics[width=0.8\textwidth]{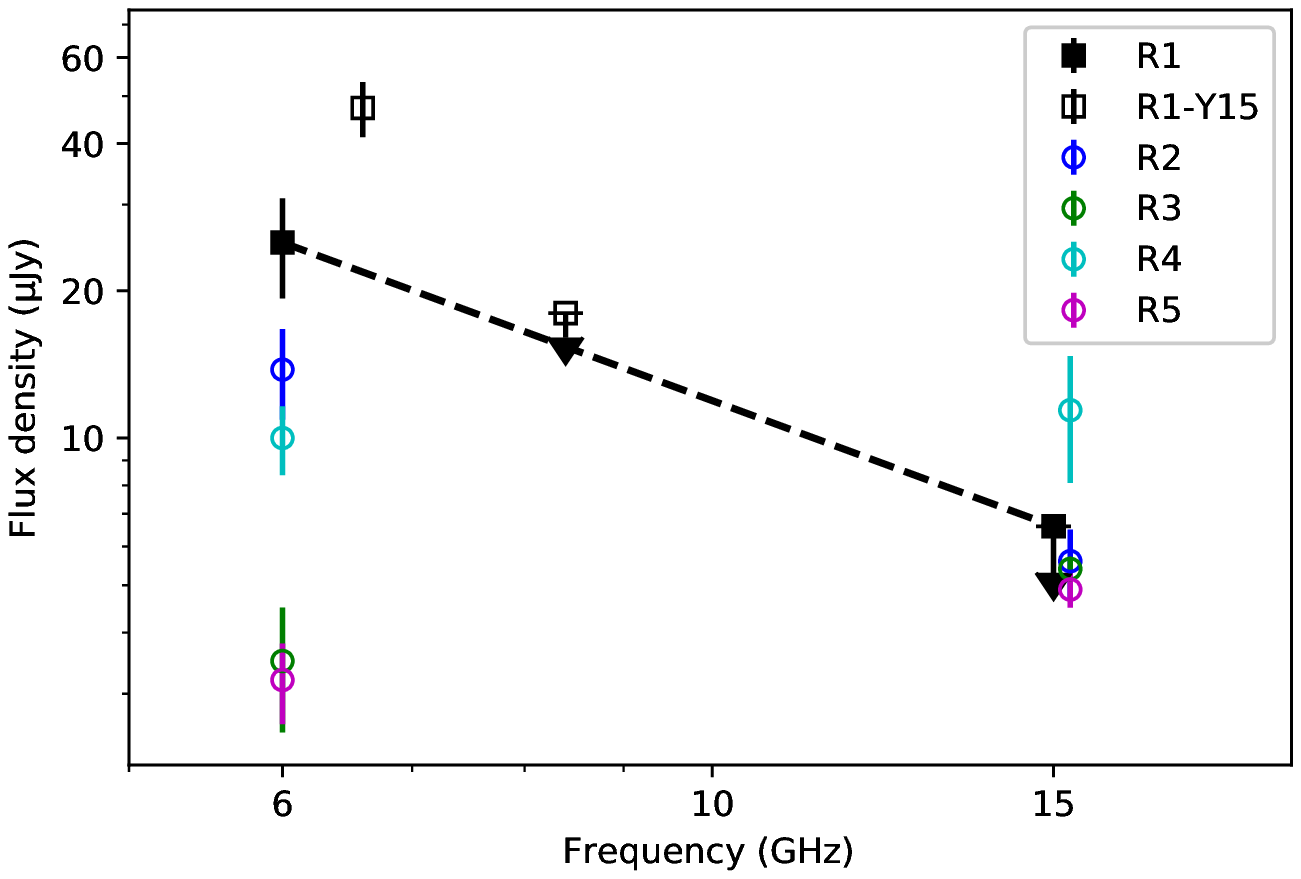}
\caption{Radio spectrum of the nuclear source (R1). Solid and open squares are the C-band measurement from A-configuration (this work) and B-configuration (Y15), respectively. 3 $\rm \sigma$ upper limits at 8.4 GHz \citep{2003ApJ...589..783H} and 15 GHz (this work) is marked by the arrows. The black dashed line linking the measurement at 6 GHz and the upper limit at 15 GHz, corresponds to a spectral index of -1.46 (i.e., regarded as an upper limit). Also shown for comparison are the four PNe which generally exhibit a much flatter spectrum (open circles, slightly shifted near 15 GHz for clarity). \label{fig:sed}} 
\end{figure}

\clearpage

\clearpage


\bibliography{M32_r1}{}
\bibliographystyle{aasjournal}

\end{document}